\documentclass[11pt]{article}
\usepackage{moriond,epsfig}

\def\La{\Lambda}
\def\Al{\overline\Lambda}
\def\XI{\Xi^{-}}
\def\aXI{\overline\Xi^+}

\def\be{\begin{equation}}
\def\ee{\end{equation}}
\def\bea{\begin{eqnarray}}
\def\eea{\end{eqnarray}}

\begin{document}
\vspace*{4cm}
\title{Results on $\Lambda$\ and $\Xi$\ production 
 in Pb-Pb collisions at 160 GeV/$c$ per nucleon
 from the NA57 experiment}

\author{Presented by G.E. Bruno for the NA57 Collaboration:
 \\
\vspace{2mm}
F.~Antinori$^{l}$,
A.~Badal{\`a}$^{g}$,
R.~Barbera$^{g}$,
A.~Belogianni$^{a}$,
A.~Bhasin$^{e}$,
I.J.~Bloodworth$^{e}$,
G.E.~Bruno$^{b}$,
S.A.~Bull$^{e}$,
R.~Caliandro$^{b}$,
M.~Campbell$^{h}$,
W.~Carena$^{h}$,
N.~Carrer$^{h}$,
R.F.~Clarke$^{e}$,
A.~Dainese$^{l}$,
A.P.~de~Haas$^{s}$,
P.C.~de~Rijke$^{s}$,
D.~Di~Bari$^{b}$,
S.~Di~Liberto$^{o}$,
R.~Divia$^{h}$,
D.~Elia$^{b}$,
D.~Evans$^{e}$,
K.~Fanebust$^{c}$,
F.~Fayazzadeh$^{k}$,
J.~Fedorisin$^{j}$,
G.A.~Feofilov$^{q}$,
R.A.~Fini$^{b}$,
J.~Ft\'a\v cnik$^{f}$,
B.~Ghidini$^{b}$,
G.~Grella$^{p}$,
H.~Helstrup$^{d}$,
M.~Henriquez$^{k}$,
A.K.~Holme$^{k}$,
A.~Jacholkowski$^{b}$,
G.T.~Jones$^{e}$,
P.~Jovanovic$^{e}$,
A.~Jusko$^{i}$,
R.~Kamermans$^{s}$,
J.B.~Kinson$^{e}$,
K.~Knudson$^{h}$,
A.A.~Kolojvari$^{q}$,
V.~Kondratiev$^{q}$,
I.~Kr\'alik$^{i}$,
A.~Kravcakova$^{j}$,
P.~Kuijer$^{s}$,
V.~Lenti$^{b}$,
R.~Lietava$^{f}$,
G.~L\o vh\o iden$^{k}$,
M.~Lupt\'ak$^{i}$,
V.~Manzari$^{b}$,
G.~Martinska$^{j}$,
M.A.~Mazzoni$^{o}$,
F.~Meddi$^{o}$,
A.~Michalon$^{r}$,
M.~Morando$^{l}$,
D.~Muigg$^{s}$,
E.~Nappi$^{b}$,
F.~Navach$^{b}$,
P.I.~Norman$^{e}$,
A.~Palmeri$^{g}$,
G.S.~Pappalardo$^{g}$,
B.~Pastir\v c\'ak$^{i}$,
J.~Pisut$^{f}$,
N.~Pisutova$^{f}$,
F.~Posa$^{b}$,
E.~Quercigh$^{l}$,
F.~Riggi$^{g}$,
D.~R\"ohrich$^{c}$,
G.~Romano$^{p}$,
K.~\v{S}afa\v{r}\'{i}k$^{h}$,
L.~\v S\'andor$^{i}$,
E.~Schillings$^{s}$,
G.~Segato$^{l}$,
M.~Sen\`e$^{m}$,
R.~Sen\`e$^{m}$,
W.~Snoeys$^{h}$,
F.~Soramel$^{l}$,
M.~Spyropoulou-Stassinaki$^{a}$,
P.~Staroba$^{n}$,
T.A.~Toulina$^{q}$,
R.~Turrisi$^{l}$,
T.S.~Tveter$^{k}$,
J.~Urb\'{a}n$^{j}$,
F.F.~Valiev$^{q}$,
A.~van~den~Brink$^{s}$,
P.~van~de~Ven$^{s}$,
P. Vande Vyvre$^{h}$,
N.~van~Eijndhoven$^{s}$,
J.~van~Hunen$^{h}$,
A.~Vascotto$^{h}$,
T.~Vik$^{k}$,
O.~Villalobos Baillie$^{e}$,
L.~Vinogradov$^{q}$,
T.~Virgili$^{p}$,
M.F.~Votruba$^{e}$,
J.~Vrl\'{a}kov\'{a}$^{j}$ and
P.~Z\'{a}vada$^{n}$
\vspace{2mm}
}
\address{$^{a}$ Physics Department, University of Athens, Athens, Greece\\
$^{b}$ Dipartimento IA di Fisica dell'Universit{\`a}
  e del Politecnico di Bari and INFN, Bari, Italy \\
$^{c}$ Fysisk Institutt, Universitetet i Bergen, Bergen, Norway\\
$^{d}$ H{\o}gskolen i Bergen, Bergen, Norway\\
$^{e}$ University of Birmingham, Birmingham, UK\\
$^{f}$ Comenius University, Bratislava, Slovakia\\
$^{g}$ University of Catania and INFN, Catania, Italy\\
$^{h}$ CERN, European Laboratory for Particle Physics, Geneva,
       Switzerland\\
$^{i}$ Institute of Experimental Physics, Slovak Academy of Science,
       Ko\v{s}ice, Slovakia\\
$^{j}$ P.J. \v{S}af\'{a}rik University, Ko\v{s}ice, Slovakia\\
$^{k}$ Fysisk Institutt, Universitetet i Oslo, Oslo, Norway\\
$^{l}$ University of Padua and INFN, Padua, Italy\\
$^{m}$ Coll\`ege de France, Paris, France\\
$^{n}$ Institute of Physics, Prague, Czech Republic\\
$^{o}$ University ``La Sapienza'' and INFN, Rome, Italy\\
$^{p}$ Dipartimento di Scienze Fisiche ``E.R. Caianiello''
       dell'Universit{\`a} and INFN, Salerno, Italy\\
$^{q}$ State University of St. Petersburg, St. Petersburg, Russia\\
$^{r}$ Institut de Recherches Subatomique, IN2P3/ULP, Strasbourg, France\\
$^{s}$ Utrecht University and NIKHEF, Utrecht, The Netherlands
}

\maketitle

\abstracts{The NA57 experiment has been designed to study the onset of enhanced 
production of multi-strange baryons and anti-baryons 
in Pb-Pb collisions with respect 
to p-Be collisions. Such enhancement, first observed by experiment WA97, 
is considered as evidence for a phase transition 
to a new state of matter -- the Quark Gluon Plasma (QGP).
In this paper, we report results on $\Lambda$\ and $\Xi$\  hyperon production  
for about the 60\% most 
central Pb-Pb collisions at 160 GeV/$c$ per nucleon beam momentum.}

\setcounter{figure}{0}

\section{Introduction}
The WA97 experiment at the CERN SPS has observed  
enhancements of 
strange and multi-strange baryon and antibaryon  
in Pb-Pb collisions 
with respect to p-Be collisions at 160 $A$\ GeV/$c$~\cite{Ant99}. 
The enhancement increases with the strangeness content of the hyperon. 
Such a behaviour has been predicted as a signature of the QGP phase 
transition~\cite{RafMul82}.

\noindent
The aim of the NA57 experiment, 
which continues and extends the measurements of WA97,  
is to study the dependence of the enhancement 
{\em (i)} on the interaction volume and {\em (ii)}  on the collision 
energy per incoming nucleon~\cite{NA57prop}. 
For the first purpose, the  
NA57 centrality range has been extended down  
to more peripheral collisions with respect to WA97;  
for the second,  the experiment has collected data using both 160 and 
40 $A$\ GeV/$c$\ beams a the CERN SPS. 

\section{The NA57 experiment}
The NA57 apparatus has been described in details elsewhere~\cite{Virgili01}. 
The tracking device consists  
of a telescope made of 13 planes of silicon pixel detectors, 
each of them with a sensitive  
area of about 5 $\times$ 5 cm$^2$. The $30 cm$ long telescope  
is placed inside a   
1.4 T magnetic field, 60 cm downstream of the lead target, inclined 
with respect to the beam line and pointing to the target so as 
to accept particles produced at central rapidity. 

\noindent
Strange and multi-strange hyperons are identified by reconstructing their weak 
decays into final states containing only charged particles, 
e.g.: $\Xi^-$ $\rightarrow$ $\Lambda\pi^-$, with 
$\Lambda$ $\rightarrow$ $\pi^-{p}$.

\noindent
The centrality of the Pb-Pb collisions is determined by analyzing the charged particle
multiplicity measured in the pseudorapidity
interval $2<\eta<4$\ by two stations of silicon strip detectors. 
An array of scintillator counters, placed  10 cm downstream of the target, 
provides a signal to trigger on the centrality of the collisions. 
The most central 60\% of the total inelastic cross section has been selected in 
Pb-Pb interactions at 160 $A$ GeV/$c$.

\noindent
To study the strange baryon and anti-baryon production at a lower center of mass energy, 
the experiment has also collected data at 40 $A$ GeV/$c$ beam momentum on both Pb-Pb and 
p-Be collisions. 
Reference data on p-Be and p-Pb at 160 GeV/$c$\ are available from the WA97 measurements. 

\noindent 
The results on $\Xi^-$, $\overline\Xi^+$, $\La$\ and $\Al$\  particles 
reported in this paper have been obtained analyzing the Pb-Pb data sample at 
160 $A$ GeV/$c$\ collected in 1998. 
A second sample of Pb-Pb data at 160 $A$ GeV/$c$\ has been collected in year 2000: the 1998 
and 2000 data sets together will allow to study the rarer $\Omega$\ signal and to double 
the present statistics for $\Xi^-$\ and $\overline\Xi^+$. 

\section{Data analysis and results}
The hyperon signals are extracted on the basis of 
kinematical selections with the same method used in the WA97 experiment~\cite{ReconInWA97}.

\begin{figure}[t]
\centering
  \includegraphics[clip,scale=0.30]{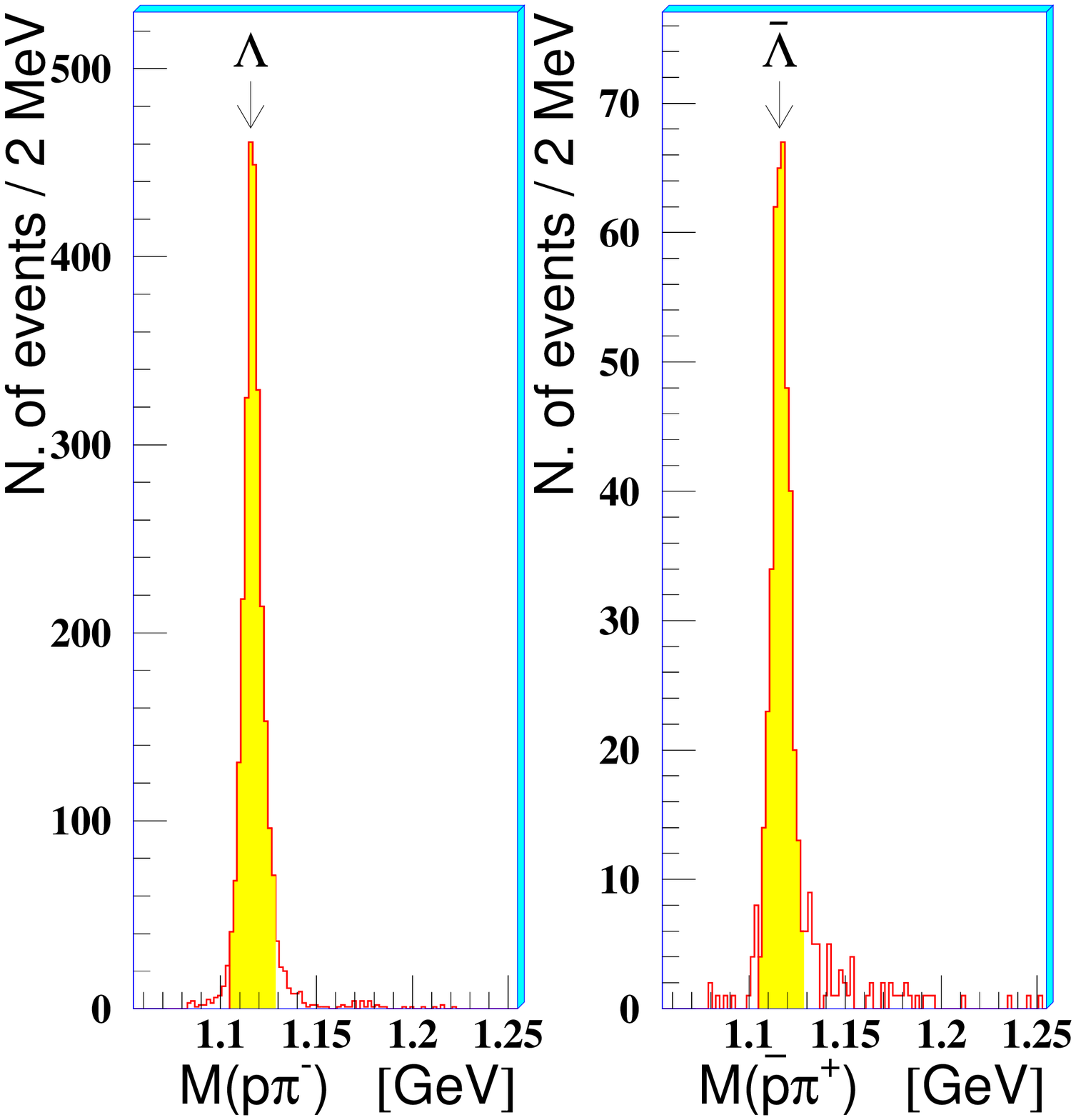}
    \hspace{0.3cm}
   \includegraphics[clip,scale=0.30]{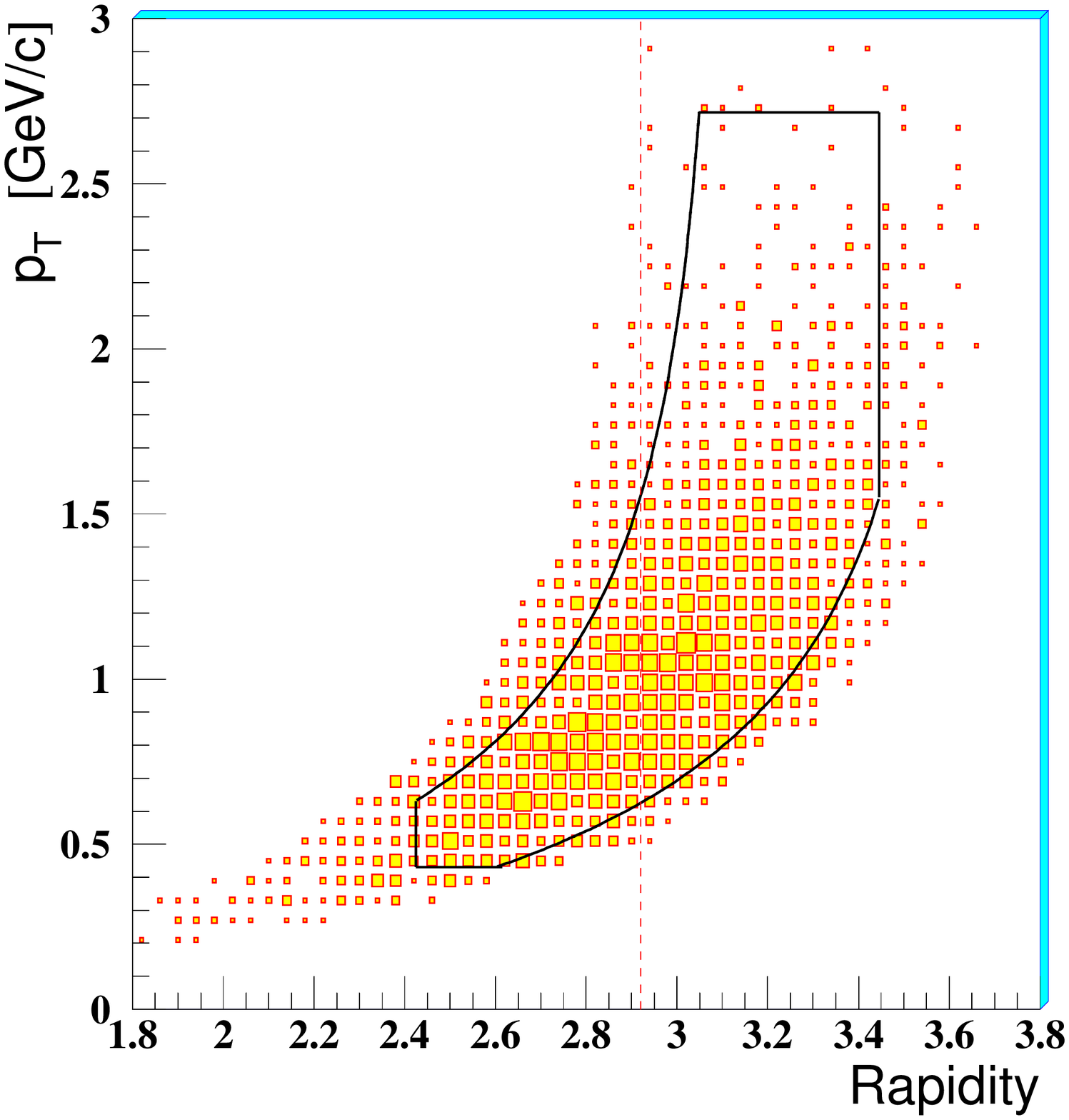} \\
   \includegraphics[clip,scale=0.30]{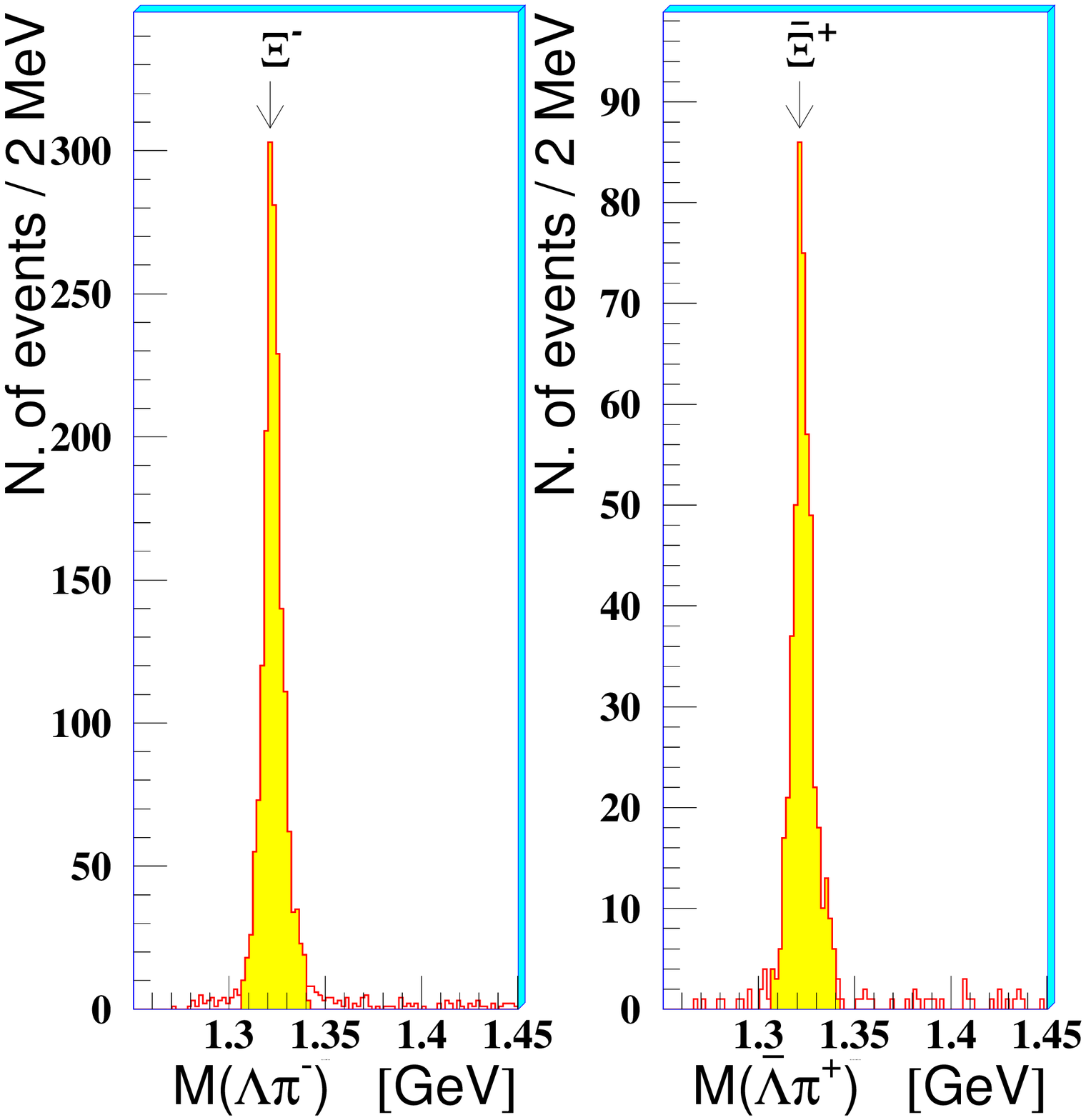}
       \hspace{0.3cm}
    \includegraphics[clip,scale=0.30]{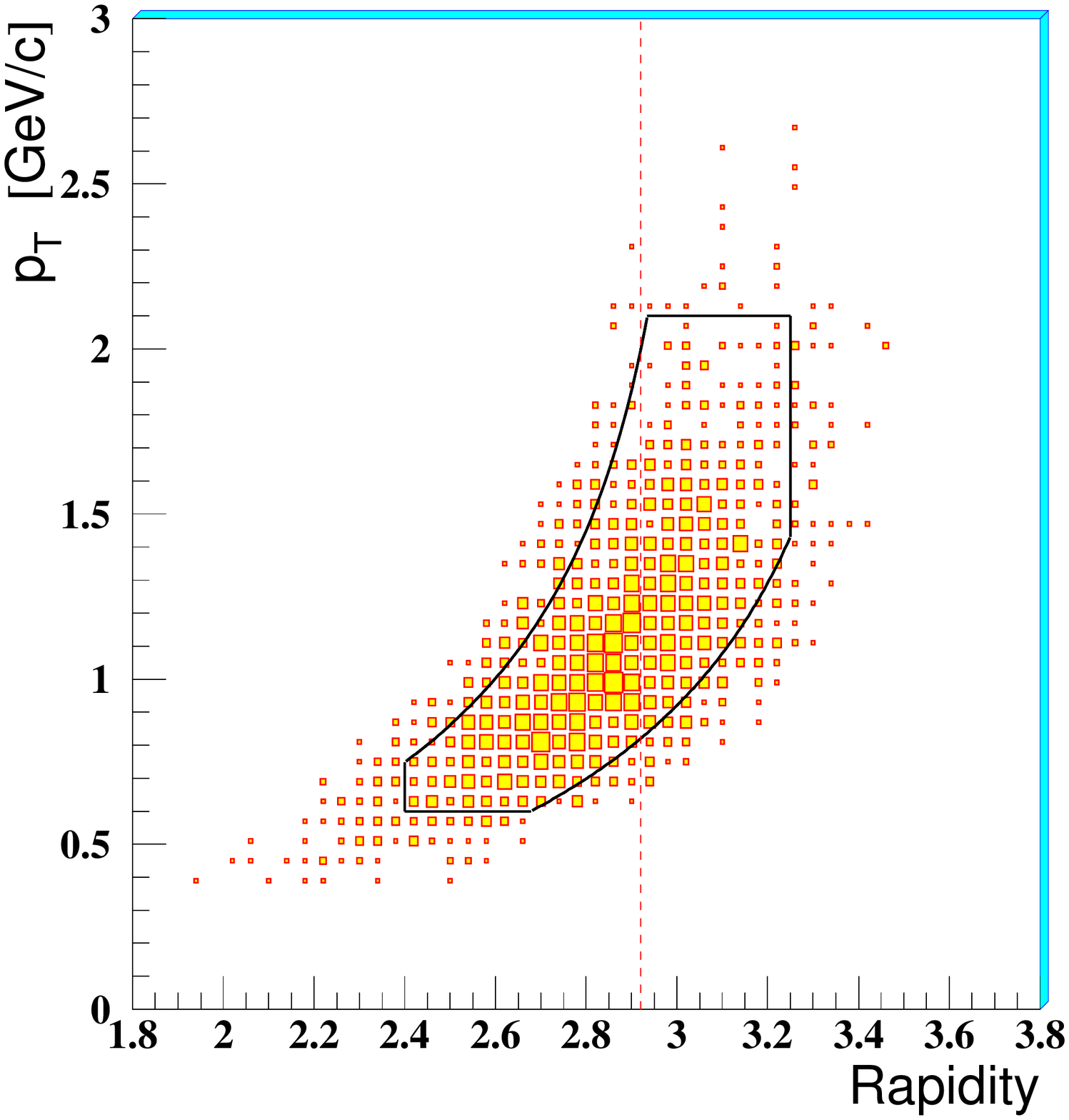}
\caption{Proton-pion ({\bf top}) and lambda-pion ({\bf bottom}) 
invariant mass distributions ({\bf left}) and
corresponding p$_T$ versus rapidity distributions ({\bf right}).  
86\% (79\%) of the reconstructed $\La$s\  ($\Xi$s) falls 
within the selected fiducial window (enclosed area).
\label{fig:Lambdasig}}
\end{figure}

\noindent
Fig.\ref{fig:Lambdasig} shows the proton-pion and the lambda-pion 
invariant mass
distributions after all the selection cuts and the
corresponding kinematic windows (enclosed areas) selected
in the y-p$_T$ distributions of the reconstructed 
particles. 
In the $y$-p$_T$ distributions the central rapidity is highlighted
with a dashed line on the acceptance window. 

\begin{figure}[t]
\centering
  \includegraphics[clip,scale=0.30]{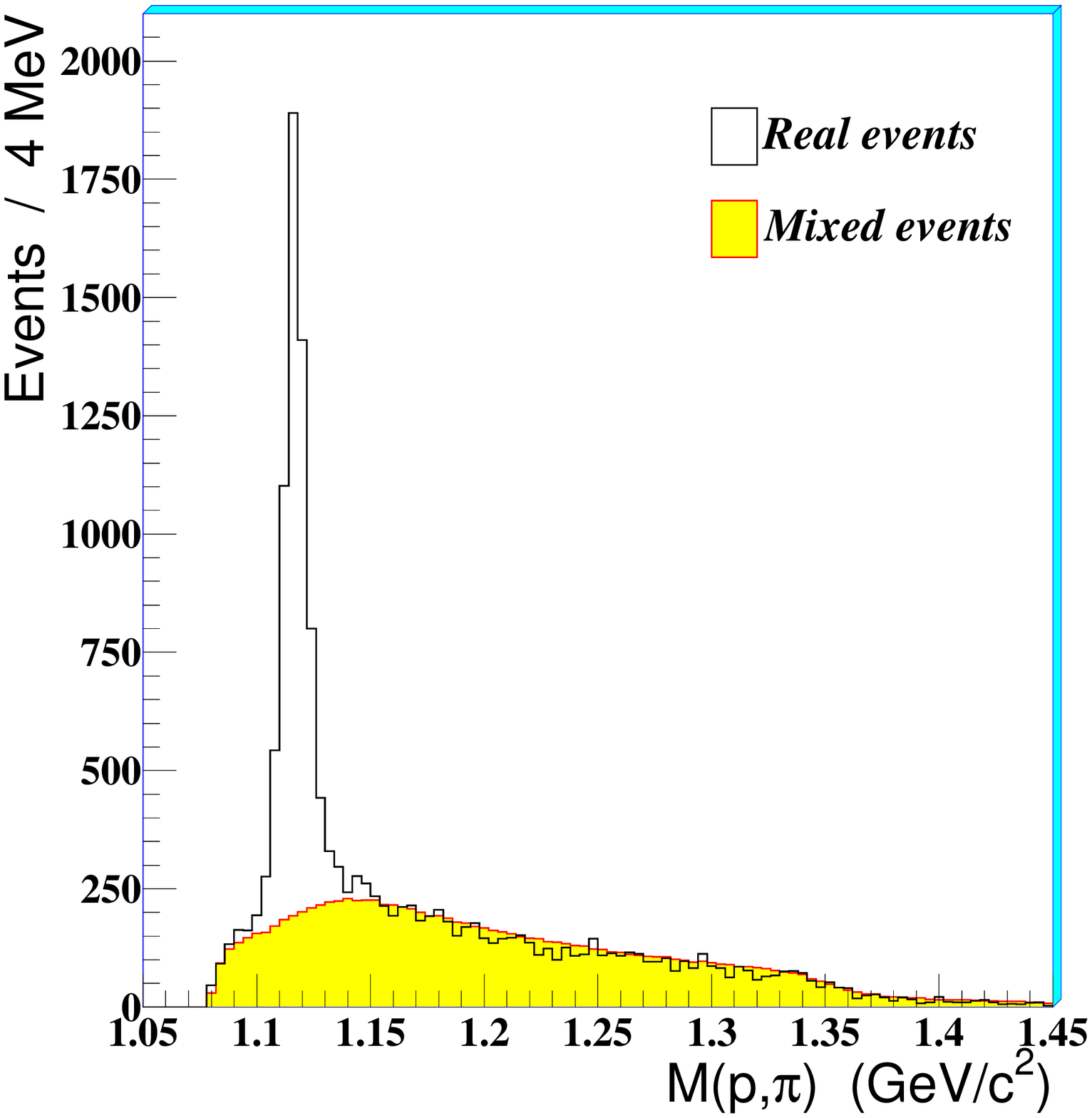}
    \hspace{0.3cm}
   \includegraphics[clip,scale=0.30]{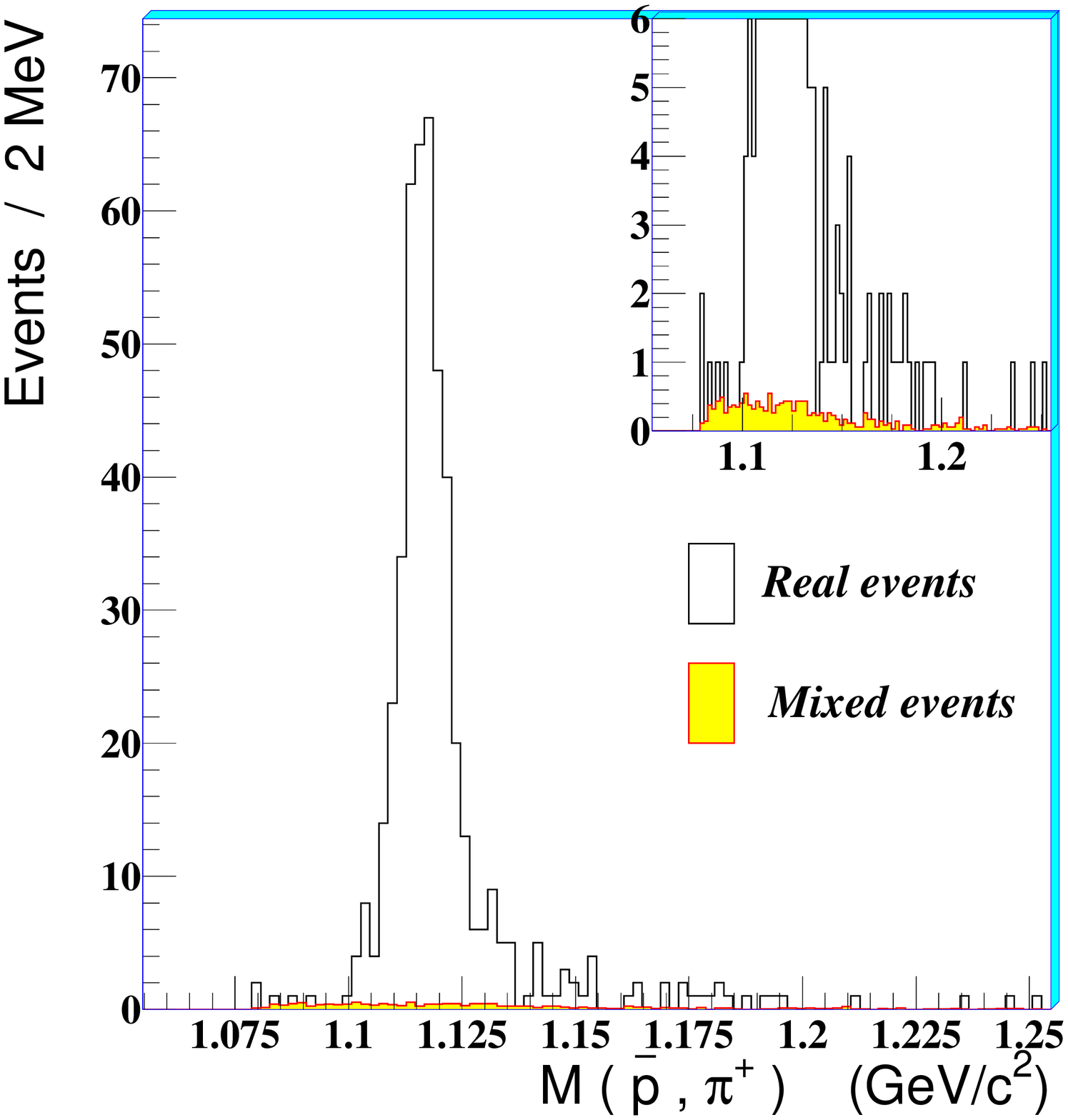}
\caption{Comparison between real and mixed events for the 
$p\pi$ invariant mass distribution before 
analysis cuts (left panel, both $\La$ and $\Al$) 
and after analysis cuts (right panel, $\Al$\ particles only). 
The insert in the right panel is a zoom on the vertical axis.
\label{fig:Mixing}}
\end{figure}

\noindent
The mass spectra for $\La$\ and $\Xi$\ are centered at the nominal value 
and 
the FWHM is about 5 MeV and 8 MeV, respectively. 
All the signals show very low background. Nevertheless, in order to study the 
shape and the total amount of the residual combinatorial background,  
we have performed a study  
using the method of {\em event mixing}.
Fake $\La$s are built 
by pairing all the negative charged particles from one event with all the positive ones 
from a different event, selecting events which are close in multiplicity.  
Then the fake  $\La$s from mixed events are reconstructed as the real ones.   
With this method any signal from a real neutral particle decaying in two 
oppositely charged particles (e.g. $\Lambda$ $\rightarrow$ $\pi {p}$, 
$K_S^0$ $\rightarrow$ $\pi^- \pi^+$, $\gamma$\ conversion)
is 
removed and what remains is the combinatorial background. 
The absolute normalization is based on the number  
of pairs of oppositely charged particles in real and mixed events.  
The mixed event sample describes very well the residual background in the physical 
distributions from real events where no signal is expected. As an example, 
Fig.~\ref{fig:Mixing} shows the $p\pi$\ invariant mass distribution for  
real and mixed events {\em before} (left) and {\em after} (right) 
the application of the analysis cuts.  
The agreement in these distributions and in others, not shown here 
(e.g. the closest distance in space between the extrapolated 
$\pi$\ and $p$\ tracks coming from the decay, 
the $\pi$\ and $p$\ impact parameters\footnote{The impact parameter is 
defined as the distance (perpendicular to the beam direction) between the 
extrapolated lines of flight of the decay particles and the primary vertex position.}, etc.), 
in regions far from any physical signal, gives confidence in the method. 
The total amount of combinatorial background is estimated to be 
about 0.3\% for $\La$\ and 1.0\% for $\Al$\ and therefore can be 
safely neglected. 
\noindent
The estimate of the total $\Xi$\ background, 
evaluated with a similar technique,  is less than 4\%. 

\noindent
In order to correct for acceptance and efficiency losses, a weight is calculated
for each particle using the following procedure\cite{Antinori2000}: 
\begin{itemize}
\item For each observed $\La$ ($\Xi$), about $35,000$\ ($250,000$) 
 Monte Carlo particles are 
 generated  with the measured transverse momentum and rapidity and with 
 random azimuthal angle;  
\item the Monte Carlo particles are traced through the apparatus by 
 GEANT~\cite{GEANT} allowing them to decay according to their proper 
 life-times  and random internal decay angles;  
\item the hits of the Monte Carlo tracks are embedded into real events to 
 account for background tracks and electronic noise;  
\item these events are reconstructed and processed 
 by the same analysis programs used for real 
 data and the weight is calculated as the ratio of the number of generated 
 Monte Carlo particles to the number of particles selected by the analysis 
 programs. 
\end{itemize}
\noindent
All the reconstructed $\XI$s and $\aXI$s from the 1998 data sample have been 
corrected with the above method;  
for the more abundant $\La$s and $\Al$s only a fraction of the available statistics,  
respectively $1/200$\ and $1/25$, has been individually weighted  
due to 
the high CPU-time expenses required to apply this method.  

\noindent
As a measure of centrality we use the number of wounded nucleons, 
i.e. the nucleons which take part in the initial collisions~\cite{Glauber}. 
The multiplicity distribution is divided into five centrality classes 
($0 ,I,II,III,IV$)~\footnote{The labels I-IV were introduced by WA97.
The NA57 most peripheral bin is indicated with symbol 0 to keep the
labelling of WA97 for the other centrality bins.},  
class $0$\ being the most peripheral 
($<N_{wound}>=62$) and 
class $IV$\ the most central ($<N_{wound}>=349$). 
The method to calculate the number of wounded 
nucleons is described elsewhere~\cite{Carrer99}.  

\noindent
The transverse mass distributions have been parametrized as:
\begin{equation}
\frac{d^2N}{dm_T dy}=A \hspace{1mm} m_T \exp\left(-\frac{m_T}{T}\right)
\label{eq:mtfit}
\end{equation}
\noindent 
and the inverse slopes $T$\ 
have been extracted by means of a maximum likelihood fit method. 
In Table~\ref{tab:InvSlopes} the inverse slopes for $\XI$ and $\aXI$ 
and the preliminary results for $\La$ and  $\Al$ 
are presented. 
\begin{table}[h]
\caption{Inverse slopes $T$\ (MeV) in the full centrality range ({\bf 0-IV}) 
 and in different centrality classes. 
\label{tab:InvSlopes}} 
\vspace{0.4cm}
\begin{center}
\begin{tabular}{|c|c||c|c|c|c|c|}
\hline
  & {\bf 0-IV} &
{\bf 0} & {\bf I} &
{\bf II} & {\bf III} 
& {\bf IV}   \\ \hline
{\bf $\La$} & {284 $\pm$\ 6} &
{258 $\pm$\ 19} & {261 $\pm$\ 11} &
{276 $\pm$\ 11} & {313 $\pm$\ 13} & {310 $\pm$\ 15} \\
{\bf $\Al$} & {287 $\pm$\ 6} &
{274 $\pm$\ 18} & {258 $\pm$\ 10} &
{279 $\pm$\ 10} & {307 $\pm$\ 13} & {309 $\pm$\ 14} \\
\cline{3-7}
{\bf $\Xi^-$} & {303 $\pm$\ 11} & 
\multicolumn{5}{|c}{} \\
{\bf $\bar{\Xi^+}$} & {321 $\pm$\ 23} &
\multicolumn{5}{|c}{} \\ 
\cline{1-2}
\end{tabular}
\end{center}
\end{table}
\noindent 
The values are found to be compatible 
for particles and their anti-particles, and with the WA97 ones. 
We observe a slight increase of $\La$ and  $\Al$ slopes 
with centrality, while no significant variation is observed~\cite{carrerqm01} 
for $\XI$ and $\aXI$ within the present statistics.

\noindent
By extrapolating Eq.~\ref{eq:mtfit} to p$_T$ = 0 and integrating 
over one unit of rapidity,   
the particle yields over the whole p$_T$ range 
are computed according to the following expression 
\begin{equation}
Yields=\int_{m}^{\infty} {\rm d}m_{T} \int_{y_{cm}-0.5}^{y_{cm}+0.5} {\rm d}y  
  \frac{d^2N}{{\rm d}m_{T} {\rm d}y}
\label{eq:yield}
\end{equation}
where $m$ is the rest mass of the particle.

\begin{figure}[ht]
\centering
  \includegraphics[clip,scale=0.52]{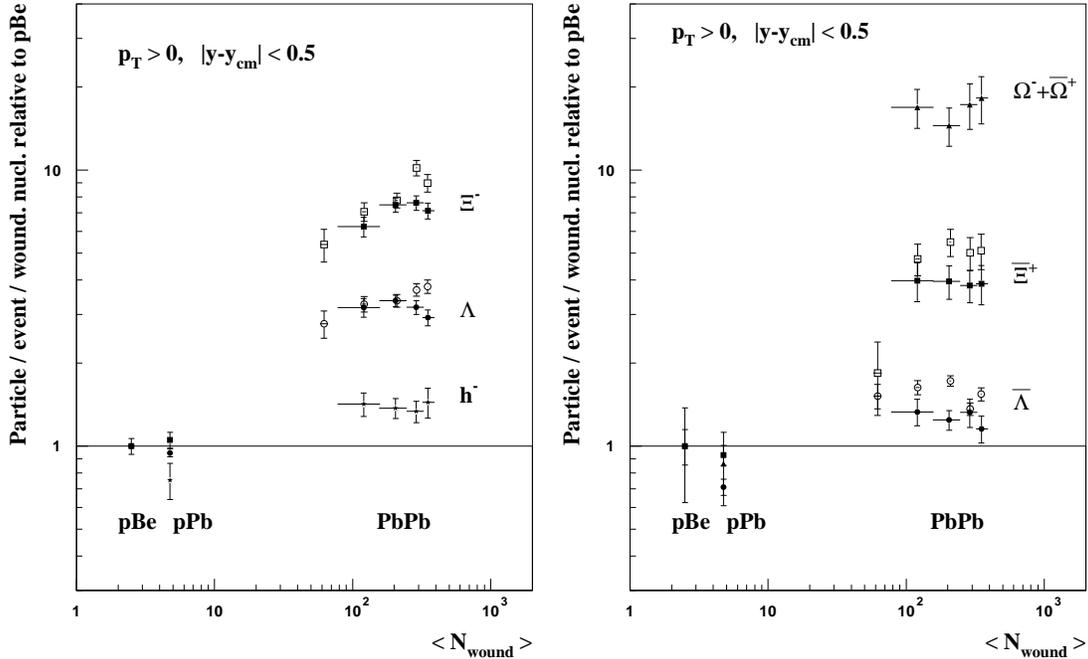}
\caption{Yields per wounded nucleon relative to p-Be
from WA97 for all measured particles (closed symbols) 
and from NA57 for $\Xi^-$, $\overline\Xi^+$, $\La$\ and $\Al$\ 
(open symbols). NA57 $\La$\ and $\Al$\ yields are preliminary. 
\label{fig:partyie}} 
\end{figure}

\noindent
In Fig.\ref{fig:partyie} the NA57 yields per participant in Pb-Pb at 160 $A$ GeV/$c$\  
relative to the p-Be ones from WA97 are shown (open symbols) 
for $\Xi^-$, $\overline\Xi^+$, $\La$\ and $\Al$\ 
in the five centrality classes, i.e. as a function of the centrality of the collision. 
\noindent 
The WA97 results are also reported in Fig.\ref{fig:partyie}  using
closed symbols.  
NA57 results confirm that the strange particle yields per participant are enhanced 
in Pb-Pb collisions with respect to p-A reference collisions.  
In the common centrality range the NA57 yields tend to be larger  
than WA97 by up to 20-30\% for $\aXI$\ and $\Al$.  
We have performed extended checks on the 
correction procedure applied for the $\Xi$~\cite{carrerqm01},  
in particular for these variables exploited in the kinematic selection 
of the particles: the outcome gives us  confidence in the NA57 
results~\cite{Kristin}.  
Similar checks are ongoing for the $\La$ hyperon. 
\noindent
$\XI$, $\aXI$ and $\La$ yields per wounded nucleon drop when going 
from $<N_{wound}>=121$ to $<N_{wound}>=62$, i.e. from class I to class 0; 
the maximum effect is observed for the $\aXI$ particle and it 
corresponds to a 3.5 $\sigma$ effect.  
This drop might indicate the onset of the QGP phase transition.  

\section{Conclusions and outlook}

Results from NA57 on $\La$\ and $\Xi$ production in Pb-Pb collisions at 160 $A$ GeV/$c$ 
have been reported. The $\XI$,  $\aXI$ and $\La$ yields per wounded nucleon in the 
most peripheral bin drop significantly (up to a 3.5 $\sigma$ effect for $\aXI$) 
which might indicate the onset of the QGP phase transition. 
The ongoing data analysis will soon provide results on the rare $\Omega$\ particle 
production at 160  $A$\ GeV/$c$\  
and the full pattern of strange particle enhancements at 40 $A$\ GeV/$c$.

\end{document}